# Mesoscopic density-operator in a uniform magnetic field

**A. Abdellaoui**[*1,2], **and F. Benamira**[1]

[1]Laboratoire de Physique Théorique, Département de Physique, Faculté des Sciences, Université Mentouri-Constantine, Constantine, Algeria
[2]Département de Physique, Centre Universitaire Larbi-Ben-M'Hidi, Oum-el-Bouaghi 04000, Algeria

Abstract. A self-consistent linear-response theory for a two-lead mesoscopic structure in Landauer's viewpoint for transport is developed. A density operator relevant in this viewpoint, in a uniform magnetic field is derived. It is shown that this operator differs from the one given in Kubo's viewpoint by a term that represents the self-consistent effects. Hence, a special emphasis is devoted to the diagonal master equation where an exact solution is obtained in the framework of elastic scattering theory.

## 1 Introduction

The idea of relating electronic transport in mesoscopic structures to their scattering properties as originally proposed by Landauer [1], has permitted a rapid growth of the quantitative study of mesoscopic phenomena about fifty years ago. There is a fundamental difference between two viewpoints of transport in mesoscopic systems. Kubo's viewpoint (KVP) considers the applied voltages as the agent causing current-flow in the sample, whereas Landauer's viewpoint (LVP) views current-flow as the causal agent and the induced potential inside the sample as the response. This difference appears quantitatively in a conductance measurement. Indeed, if the voltage difference is measured between the reservoirs (KVP), the corresponding conductance is linear on the transmission coefficient of the sample [2]. However, if the voltage difference is measured with regards to the induced potential, the conductance formula is a non-linear function on the transmission coefficient [1]. It is now well established that contacts between the reservoirs and the sample are at the origin of the difference between the two formulas [3].

In this work, by performing a self-consistent linear-response theory (LRT), we show that the fundamental difference between KVP and LVP manifests within the density-operator.

## 2 Density-operator in the viewpoint of Landauer

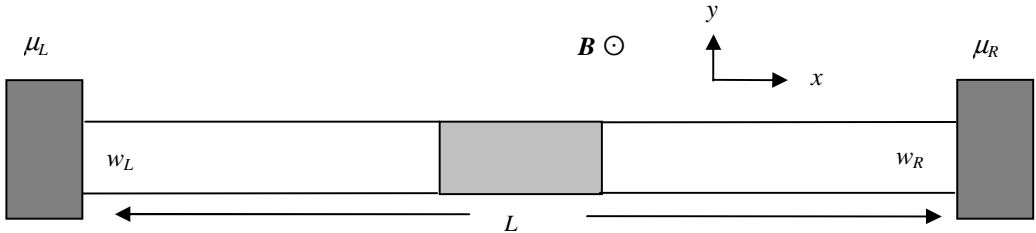

**Fig. 1** Mesoscopic sample in a perpendicular magnetic field

---

[*] e-mail: abdellaouia@wissal.dz

Consider a mesoscopic scatterer connected via identical perfect leads to two large particle-reservoirs that are meant to act as current source and sink. The sample (the scatterer + leads) is immersed in a uniform magnetic field $\boldsymbol{B}=B\hat{z}$ (Fig.1) and throughout it the transport is completely coherent.

We assume that at $t<0$ the system was in thermal equilibrium with an equilibrium density operator $\rho_0$ corresponding to the case where the reservoirs are maintained at the same chemical potential $\mu_R$. The system is then spontaneously driven out of equilibrium at $t=0$ by changing the chemical potential of the left reservoir from $\mu_R$ to $\mu_L$ so that $\mu_L-\mu_R$ is small enough to ensure a linear-transport regime. An electric current begins thus to circulate in the sample and reaches a steady-state (SS) value after a long time.

In LVP, upon the introduction of the scattering potential, an additional transport field emerges into the sample as a response to the current-flow. Due to charges accumulation on both sides of the barrier, a potential built-up across it and inside the perfect leads. This potential, thereafter denoted $w(\boldsymbol{r})$, allows for a self-consistent screening of the piled-up charges [4]. In the SS-regime, $w(\boldsymbol{r})$ will get constant asymptotic values $w_L$ and $w_R$ in the vicinity of the reservoirs with $w_L \neq \mu_L$ and $w_R \neq \mu_R$, respectively (See Fig.1). The full one-particle Hamiltonian for non-interacting carriers is given by

$$H' = H + w(\boldsymbol{r}) = H_0 + V(\boldsymbol{r}) + w(\boldsymbol{r}), \tag{1}$$

with $w(\boldsymbol{r})$ is considered as the external perturbation, $V(\boldsymbol{r})$ is the electron-impurity interaction, and $H_0$ is the free confined electrons Hamiltonian. In the effective mass approximation $H_0 = 1/2m^*(\boldsymbol{p}-e\boldsymbol{A})^2 + V_c(\boldsymbol{r})$, where $V_c(\boldsymbol{r})$ is the confinement potential in the $y$ and $z$ directions. Taking the Landau gauge ($\boldsymbol{A}=-By\boldsymbol{i}$), the eigenstates of $H_0$ are a product of plane waves in the longitudinal direction $x$ and of confined functions in the transverse directions. The eigenstates of $H_0$ and $H$, denoted by $|\varphi_\xi\rangle$ and $|\psi_\xi\rangle$ respectively with $\xi$ denotes a complete set of quantum numbers necessary to describe the corresponding dynamics, form two complete orthonormal bases.

Due to current-flow, the driven one particle density-operator $\rho(t) = \rho(t) - \rho_0$ that represents the deviation of the density operator $\rho(t)$ from its equilibrium value $\rho_0$, satisfies in LR-approximation the given one-particle Von-Neumann equation

$$\frac{\partial}{\partial t}\rho(t) + i\ell\rho(t) = -\frac{i}{\hbar}[w(\boldsymbol{r}), \rho_0]. \tag{2}$$

$\ell$ is the Liouvillian superoperator associated with $H$, i.e. $\ell\boldsymbol{\cdot} = \hbar^{-1}[H, \boldsymbol{\cdot}]$, and $\rho(0)=0$.

By taking Laplace-Carson transform, defined by $\hat{g}(s) = s\int_0^\infty dt\, e^{-st} g(t)$, on both sides of Eq. (2), we obtain

$$s\hat{\rho}(s) + i\ell\hat{\rho}(s) = -\frac{i}{\hbar}[w(\boldsymbol{r}), \rho_0]. \tag{3}$$

This Laplace-Carson technique is mainly equivalent to the technique proposed in the original work of Kubo [5] as well as to the alternative approach given by Lax [6]. Its advantage is that the SS-solution $\rho$ may be easily obtained from $lim_{s\to 0^+} \hat{\rho}(s) = lim_{t\to\infty}\rho(t) = \rho$. However, we show that the $lim_{s\to 0^+}$ on Eq. (3) has to be taken carefully according to every physical situation. In LVP we have to take the $lim_{s\to 0^+}$ in Eq. (3) at the beginning and before solving it, whereas in KVP, Eq. (3) must be solved first and the $lim_{s\to 0^+}$ taken at the end of calculation. Lax [6] has shown in a former work that $s\hat{\rho}(s)$ expresses the effect of the surroundings. In our case, it corresponds to the effect of the reservoirs (contacts) on the transport carriers. Indeed, in KVP, when the reservoirs are included in the measure of the voltage difference, $\hat{\rho}(s)$ presents a diagonal singularity [7]. Consequently, the $lim_{s\to 0^+}$ in Eq. (3) must be taken at the end of the calculations. However, in LVP, voltage measurement excludes the contacts. In this case $\hat{\rho}(s)$ presents no diagonal singularity so that the $lim_{s\to 0^+}$ may be taken at the beginning.

### 1.1 Kubo's density-operator

In KVP, the SS-density operator reads then

$$\rho^K = -\frac{i}{\hbar}\lim_{s\to 0^+}\frac{1}{s+i\ell}[w, \rho_0]. \tag{4}$$

Note that in a former work [8], the authors considered that the asymptotic values of the induced potential, $w_L$ and $w_R$, mimic the chemical potentials of the adjacent reservoirs $\mu_L$ and $\mu_R$. Moreover, ignoring self-consistent effects, the authors obtained the Landauer-Büttiker conductance formula [2].

### 1.2 Landauer's density-operator

In LVP, Eq. (3) has to be solved without taking into account the effect of the contacts. This is achieved by excluding the diagonal part of $s\hat{\rho}(s)$ from Eq. (3). For this, we will appeal to the projection formalism. Using the projection superoperators $P$ and $Q=1-P$, the density operator $\hat{\rho}(s)$ is expressed on the basis of $H_0$ like: $\hat{\rho}(s) = \hat{\rho}_d(s) + \hat{\rho}_{nd}(s)$, where $\hat{\rho}_d(s) = P\hat{\rho}(s) = \sum_\xi |\varphi_\xi\rangle\langle\varphi_\xi|\langle\varphi_\xi|\hat{\rho}(s)|\varphi_\xi\rangle$ and $\hat{\rho}_{nd}(s) = Q\hat{\rho}(s) = \sum_{\xi\neq\xi'} |\varphi_\xi\rangle\langle\varphi_{\xi'}|\langle\varphi_\xi|\hat{\rho}(s)|\varphi_{\xi'}\rangle$ represent the diagonal and non-diagonal parts of the density operator, respectively. Thus, acting by $P$ and $Q$ on Eq. (3), we obtain

$$s\hat{\rho}_d(s) + iP\ell^1\hat{\rho}_{nd}(s) = \frac{-i}{\hbar}P[w,\rho_0], \tag{5a}$$

$$s\hat{\rho}_{nd}(s) + i(\ell^0 + Q\ell^1)\hat{\rho}_{nd}(s) = -i\ell^1\hat{\rho}_d(s) - \frac{i}{\hbar}Q[w,\rho_0], \tag{5b}$$

two-coupled equations for $\hat{\rho}_d(s)$ and $\hat{\rho}_{nd}(s)$, where $\ell^0$ and $\ell^1$ are the Liouvillian superoperators; $\ell^0. = \hbar^{-1}[H_0,.]$ and $\ell^1. = \hbar^{-1}[V,.]$. Till now, both equations (5a) and (5b) are equivalent to Eq. (3). To exclude the effect of the contacts from the SS-solution of (5a) and (5b), we have now to set $s\hat{\rho}_d(s) = 0$ before uncoupling these equations. Therefore, the SS-solutions yield

$$\Lambda \rho_d^L = -\frac{i}{\hbar} \lim_{s\to 0^+} P\left\{1 - i\ell^1 \frac{1}{s+i\ell}\right\}[w,\rho_0], \tag{6a}$$

$$\rho_{nd}^L = \rho^K - \lim_{s\to 0^+} \frac{i}{s+i\ell}\ell^1\rho_d^L, \tag{6b}$$

where $\Lambda = \lim_{s\to 0^+} P\ell^1 \frac{1}{s+i\ell}\ell^1 P$ is thereafter called master superoperator.

In Ref. [9] it has been shown that only the diagonal part $\rho_d^L$ given by the solution of the master equation (6a) is relevant in a two-lead mesoscopic sample in zero magnetic field. This work is devoted to the full density operator relevant in the presence of a magnetic filed.

By adding $\rho_d^L$ to the two sides of Eq. (6b), we obtain the formal density operator relevant in LVP as

$$\rho^L = \rho^K + \lim_{s\to 0^+}\left(1 - \frac{i}{s+i\ell}\ell^1\right)\rho_d^L. \tag{7}$$

To put the second term in Eq. (7) in a compact form, we use the outgoing $|\psi_\xi^+\rangle$ and ingoing $|\psi_\xi^-\rangle$ scattering states as the eigenstates of $H$. In the asymptotic regions the corresponding wave functions are simple combinations of the eigenfunctions of $H_0$: they are related via the transmission and reflection coefficients $t_{ba}^{L,R}$ and $r_{ba}^{L,R}$ [8, 9]. Indeed, a straightforward calculation yields

$$\rho^L = \rho^K + \sum_\xi |\psi_\xi^+\rangle\langle\psi_\xi^+|\rho_{\xi\xi}^L, \tag{8}$$

where $\rho_{\xi\xi}^L = \langle\varphi_\xi|\rho_d^L|\varphi_\xi\rangle$. As it appears, there is a fundamental difference between Landauer's and Kubo's density operators which resides in the term $\sum_\xi |\psi_\xi^+\rangle\langle\psi_\xi^+|\rho_{\xi\xi}^L$ that may be seen as responsible of self-consistent effects. In order to make quantitatively this term more explicit, we have to determine the matrix elements of $\rho_d^L$ from solution of the master equation (6a). Indeed, following the same steps as in Ref. [9], we show that even in the presence of a uniform magnetic field, Eq. (6a) fulfills

$$\left(\Lambda\rho_d^L\right)_{\xi\xi} = \sum_\zeta \Lambda_{\xi\xi|\zeta\zeta}\rho_{\zeta\zeta}^L = \Delta w/e\left[-f'(\varepsilon_\xi)\right]\langle\psi_\xi^-|i|\psi_\xi^-\rangle, \tag{9}$$

where $\Lambda_{\xi\xi|\zeta\zeta} = \frac{2\pi}{\hbar}\sum_\alpha \delta(\varepsilon_\xi - \varepsilon_\alpha)|\langle\varphi_\xi|V|\psi_\alpha^+\rangle|^2(\delta_{\xi\zeta} - \delta_{\alpha\zeta})$, $\Delta w = w_L - w_R$, $f'(\varepsilon)$ is the Fermi-Dirac derivative distribution function and $i$ is the current-flow operator. The quantum numbers namely $\xi$, consists of the energy $\varepsilon_\xi$, the transmitted channel $a$, and the direction of propagation $\sigma = \pm$. Carrying out the integration over $\varepsilon_\alpha$ on the left hand side of Eq. (9), this latter may be put in the following form, for fixed energy $\varepsilon_\xi = \varepsilon$,

$$\left(\Lambda\rho_d^L\right)_{a\sigma}^\varepsilon = \frac{1}{L}v_a \sum_{b\sigma'} (\Gamma)_{a\sigma|b\sigma'}^\varepsilon \left(\rho_d^L\right)_{b\sigma'}^\varepsilon, \tag{10}$$

where $L$ is the length of the sample, $v_a$ is the channel velocity and $\Gamma$ is a new matrix given by:

$(\Gamma)_{a+|b+}^\varepsilon = \delta_{ab} - |t_{ab}^L|^2$ ; $(\Gamma)_{a-|b-}^\varepsilon = \delta_{ab} - |t_{ab}^R|^2$ ; $(\Gamma)_{a-|b+}^\varepsilon = -|r_{ab}^L|^2$ ; $(\Gamma)_{a+|b-}^\varepsilon = -|r_{ab}^R|^2$.

Thus, combined together, at given energy Eqs (9) and (10) lead to

$$\left(\rho_d^L\right)_{a\sigma}^\varepsilon = \Delta w[-f'(\varepsilon)]\sum_{b\sigma'} \left(\overline{\Gamma}^{-1}\right)_{a\sigma|b\sigma'}^\varepsilon \left(\Pi^-\right)_{b\sigma'}^\varepsilon, \tag{11}$$

with $\overline{\Gamma}^{-1}$ is the inverse of $\Gamma$ [10], and $\Pi^-$ is related to the left and right transmission amplitudes by: $(\Pi^-)_{a+}^\varepsilon = (\mathbf{t}^L \mathbf{t}^{L+})_{aa}$ ; $(\Pi^-)_{a-}^\varepsilon = -(\mathbf{t}^R \mathbf{t}^{R+})_{aa}$. In the limit of weak transmission, we can see that these matrix elements are negligible. Consequently, in that limit, the self-consistent term in Eq. (8) may be neglected and the density operators $\rho^L$ and $\rho^K$ will give the same result.

## 3 Conclusion

The result presented here gives a quantitative formulation of the LVP within LRT. It provides an important step in proving the difference between LVP and KVP which manifests in the density operator itself. Therefore, our result may be used to evaluate all physical quantities whenever self-consistent effects are relevant, otherwise whenever the effect of the contacts is neglected and the induced electrochemical potential in the leads is considered. Indeed, in a conductance description, we show that while $\rho^K$ gives the Landauer-Büttiker formula [2], $\rho^L$ gives a generalized Landauer formula [10].